\documentclass[twoside,twocolumn,9pt]{article}
\usepackage{lipsum}   %to generate filler text
\usepackage{extsizes}
\usepackage[super,sort&compress,comma]{natbib} 
\usepackage[version=3]{mhchem}
\usepackage[left=1.5cm, right=1.5cm, top=1.785cm, bottom=2.0cm]{geometry}
\usepackage{balance}
\usepackage{mathptmx}
\usepackage{sectsty}
\usepackage{graphicx} 
\usepackage{lastpage}
\usepackage[format=plain,justification=justified,singlelinecheck=false,font={stretch=1.125,small,sf},labelfont=bf,labelsep=space]{caption}
\usepackage{float}
\usepackage{fancyhdr}
\usepackage{fnpos}
\usepackage[english]{babel}
\addto{\captionsenglish}{%
  
}
\usepackage{array}
\usepackage{droidsans}
\usepackage{charter}
\usepackage[T1]{fontenc}
\usepackage[usenames,dvipsnames]{xcolor}
\usepackage{setspace}
\usepackage[compact]{titlesec}
\usepackage{hyperref}
\definecolor{cream}{RGB}{222,217,201}

\begin{document}

\pagestyle{fancy}
\thispagestyle{plain}
\fancypagestyle{plain}{
%%%HEADER%%%
\renewcommand{\headrulewidth}{0pt}
}
%%%END OF HEADER%%%

%%%PAGE SETUP - Please do not change any commands within this section%%%
\makeFNbottom
\makeatletter
\renewcommand\LARGE{\@setfontsize\LARGE{15pt}{17}}
\renewcommand\Large{\@setfontsize\Large{12pt}{14}}
\renewcommand\large{\@setfontsize\large{10pt}{12}}
\renewcommand\footnotesize{\@setfontsize\footnotesize{7pt}{10}}
\makeatother

\renewcommand{\thefootnote}{\fnsymbol{footnote}}
\renewcommand\footnoterule{\vspace*{1pt}% 
\color{cream}\hrule width 3.5in height 0.4pt \color{black}\vspace*{5pt}} 
\setcounter{secnumdepth}{5}

\makeatletter 
\renewcommand\@biblabel[1]{#1}            
\renewcommand\@makefntext[1]% 
{\noindent\makebox[0pt][r]{\@thefnmark\,}#1}
\makeatother 
\renewcommand{\figurename}{\small{Fig.}~}
\sectionfont{\sffamily\Large}
\subsectionfont{\normalsize}
\subsubsectionfont{\bf}
\setstretch{1.125} %In particular, please do not alter this line.
\setlength{\skip\footins}{0.8cm}
\setlength{\footnotesep}{0.25cm}
\setlength{\jot}{10pt}
\titlespacing*{\section}{0pt}{4pt}{4pt}
\titlespacing*{\subsection}{0pt}{15pt}{1pt}
%%%END OF PAGE SETUP%%%

%%%FOOTER%%%
\fancyfoot{}
\fancyfoot[LO,RE]{\vspace{-7.1pt}\includegraphics[height=9pt]{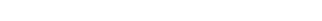}}
\fancyfoot[CO]{\vspace{-7.1pt}\hspace{13.2cm}\includegraphics{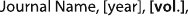}}
\fancyfoot[CE]{\vspace{-7.2pt}\hspace{-14.2cm}\includegraphics{head_foot/RF}}
\fancyfoot[RO]{\footnotesize{\sffamily{1--\pageref{LastPage} ~\textbar  \hspace{2pt}\thepage}}}
\fancyfoot[LE]{\footnotesize{\sffamily{\thepage~\textbar\hspace{3.45cm} 1--\pageref{LastPage}}}}
\fancyhead{}
\renewcommand{\headrulewidth}{0pt} 
\renewcommand{\footrulewidth}{0pt}
\setlength{\arrayrulewidth}{1pt}
\setlength{\columnsep}{6.5mm}
\setlength\bibsep{1pt}
%%%END OF FOOTER%%%

%%%FIGURE SETUP - please do not change any commands within this section%%%
\makeatletter 
\newlength{\figrulesep} 
\setlength{\figrulesep}{0.5\textfloatsep} 

\newcommand{\topfigrule}{\vspace*{-1pt}% 
\noindent{\color{cream}\rule[-\figrulesep]{\columnwidth}{1.5pt}} }

\newcommand{\botfigrule}{\vspace*{-2pt}% 
\noindent{\color{cream}\rule[\figrulesep]{\columnwidth}{1.5pt}} }

\newcommand{\dblfigrule}{\vspace*{-1pt}% 
\noindent{\color{cream}\rule[-\figrulesep]{\textwidth}{1.5pt}} }

\makeatother
%%%END OF FIGURE SETUP%%%

%%%TITLE, AUTHORS AND ABSTRACT%%%
\twocolumn[
  \begin{@twocolumnfalse}
{\includegraphics[height=30pt]{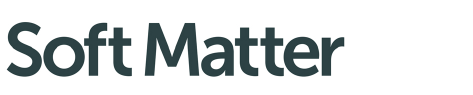}\hfill\raisebox{0pt}[0pt][0pt]{\includegraphics[height=55pt]{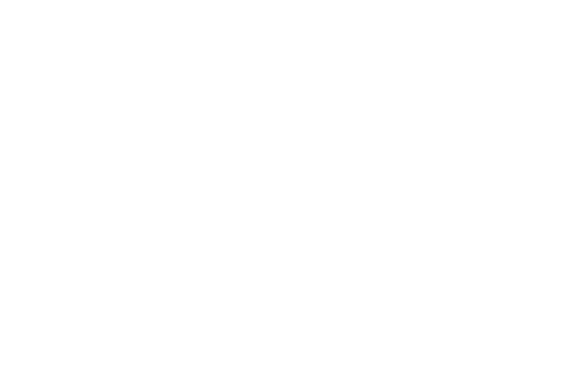}}\\[1ex]
\includegraphics[width=18.5cm]{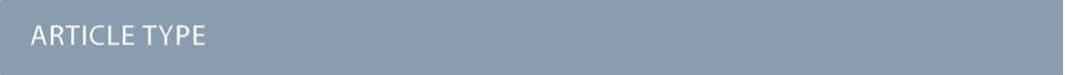}}\par
\vspace{1em}
\sffamily
\begin{tabular}{m{4.5cm} p{13.5cm} }

\includegraphics{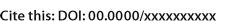} & \noindent\LARGE{\textbf{Scale-dependent sharpening of interfacial fluctuations in shape-based models of dense cellular sheets}}%Spectral analysis reveals length-scale dependent sharpening effect in Voronoi models$^\dag$}} 
\\%Article title goes here instead of the text "This is the title"
\vspace{0.3cm} & \vspace{0.3cm} \\

 & \noindent\large{Haicen Yue,\textit{$^{a}$} Charles R. Packard,\textit{$^{b}$} and Daniel M. Sussman\textit{$^{b}$}$^{\ast}$} \\%Author names go here instead of "Full name", etc.

\includegraphics{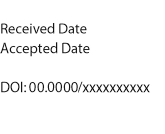} & \noindent
\normalsize{The properties of tissue interfaces -- between separate populations of cells, or between a group of cells and its environment -- has attracted intense theoretical, computational, and experimental study. Recent work on shape-based models inspired by dense epithelia have suggested a possible ``topological sharpening'' effect, by which four-fold vertices spatially coordinated along a cellular interface lead to a cusp-like restoring force acting on cells at the interface, which in turn greatly suppresses interfacial fluctuations. We revisit these interfacial fluctuations, focusing on the distinction between  short length scale reduction of interfacial fluctuations and long length scale renormalized surface tension. To do this, we implement a spectrally resolved analysis of fluctuations over extremely long simulation times. This leads to more quantitative information on the topological sharpening effect, in which the degree of sharpening depends on the length scale over which it is measured. We compare our findings with a Brownian bridge model of the interface, and close by analyzing existing experimental data in support of the role of short-length-scale topological sharpening effects in real biological systems. 
} \\

\end{tabular}

 \end{@twocolumnfalse} \vspace{0.6cm}

  ]
%%%END OF TITLE, AUTHORS AND ABSTRACT%%%

%%%FONT SETUP - please do not change any commands within this section
\renewcommand*\rmdefault{bch}\normalfont\upshape
\rmfamily
\section*{}
\vspace{-1cm}

%%%FOOTNOTES%%%

\footnotetext{\textit{$^{a}$~Department of Physics, University of Vermont, Burlington, Vermont 05405, USA; E-mail: haicen.yue@uvm.edu}}
\footnotetext{\textit{$^{b}$~Department of Physics, Emory University, Atlanta, Georgia 30322, USA }}
\footnotetext{$^{\ast}$~Email: daniel.m.sussman@emory.edu}
%Please use \dag to cite the ESI in the main text of the article.
%If you article does not have ESI please remove the the \dag symbol from the title and the footnotetext below.
%\footnotetext{\dag~Electronic Supplementary Information (ESI) available: [details of any supplementary information available should be included here]. See DOI: 10.1039/cXsm00000x/}
%additional addresses can be cited as above using the lower-case letters, c, d, e... If all authors are from the same address, no letter is required

%\footnotetext{\ddag~Additional footnotes to the title and authors can be included \textit{e.g.}\ `Present address:' or `These authors contributed equally to this work' as above using the symbols: \ddag, \textsection, and \P. Please place the appropriate symbol next to the author's name and include a \texttt{\textbackslash footnotetext} entry in the the correct place in the list.}

%%%END OF FOOTNOTES%%%

%%%MAIN TEXT%%%%
\section{Introduction}
Interfaces between populations of cells, or between cells and their environment, is pivotal for a wide variety of biological processes, ranging from embryonic development to wound healing to  tumor metastasis to lineage sorting \cite{dahmann2011boundary, yang2020leader, pawlizak2015testing,yanagida2022cell}. The stability and structure of the interface unveils information of cell and tissue mechanics, and the corresponding pathological changes in certain diseases. In perhaps the most famous model for cell-cell interfaces, the differential adhesion hypothesis (DAH) \cite{steinberg1962mechanism, steinberg2007differential}distinct cell groups are treated as immisible liquids with effective surface tension originating from different cell-cell adhesion. Other cellular mechanisms, such as cortical tension, can also contribute to the effective surface tension \cite{amack2012knowing,manning2010coaction,youssef2011quantification,krieg2008tensile,brodland2002differential}. While theoretically appealing, there have long been questions about whether the DAH and related theories are sufficient to understand the boundaries between cellular populations\cite{pawlizak2015testing,wiseman1977can,ninomiya2012cadherin,monier2010actomyosin}. At a qualitative level, it is often observed that very sharp, low-roughness boundaries exist between coexisting groups of cells, and it is not clear at a quantitative level whether the molecular mechanisms of generating adhesion and tissue surface tension are sufficent to account for this observed boundaries \cite{amack2012knowing}.

Recently it has been observed that a popular simplified class of models of dense epithelial tissues  -- vertex and Voronoi models of confluent cells \cite{honda1983geometrical,bi2016motility,alt2017vertex} -- might possess an unusual mechanism that could lead to sharp interfaces at very little energetic cost. We describe these geometrical, shape-based models in more detail below, but the essential idea is that at interfaces they might possess a non-analytic, cusp-like potential mediated by the coordination of four-fold vertices (corresponding to either highly-coordinated vertices \emph{or} to very short edges in real cellular systems) \cite{sussman2018soft}. This cusp-like potential for cells at the interface can give rise to an apparent effective surface tension larger than the microscopic surface energy term -- this can surpress interfacial fluctuations, leading to spatial registry of cells in which cell positions are highly correllated across the interface, yet still lead to relatively compliant behavior with respect to mechanical perturbations \cite{sussman2018soft, sahu2021geometric, lawson2024differences}.

This proposed underlying mechanism involves a cusp-like non-analyticity on the scale of small numbers of cells affecting the strength of the effective surface tension on a much larger length scale. Much like any coarse-grained analyis of a specific system \cite{kardar2007statistical}, from this mechanism one would expect that on short length scales the interface should be dominated by the specific microscopic physics governing the model, and at very large length scales these specific features should eventually average out into an interface governed by a standard surface energy but with a renormalized value of the surface tension. However, it is difficult to analytically compute this renormalization, and existing numerical studies give little quantitative indication of how strong the sharpening effect is at different length scales. 

Additionally, although previous studies showed that both Voronoi and vertex models have cusp-like potentials near four-fold vertices, the specific mechanisms are different and result in different interfacial behavior from the perspective of microscopic restoring forces\cite{lawson2024differences}. From the experimental perspective, although four-fold vertices are observed in biological systems, there is as-yet no evidence directly supporting the hypothesis that the effect observed in these computational  models plays a role in sharpening the boundary in real systems. This is in part due to the fact that when only the average width of the experimental interfaces is measured this allows one to estimate an \emph{effective} surface tension; it is difficult to further estimate which part of the effective surface tension originates from the explicit differences of cell-cell adhesion and cortical tension, and which which part is due to the proposed topological sharpening effect.     

In this paper, we revisit the interfacial fluctuations of these shape-based models, focusing on this distinction between a short length scale reduction of interfacial fluctuations and long length scale renormalized surface tension. We do this by studying both the average width of cell-cell interfaces, and also implementing a spectrally resolved analysis of interfacial fluctuations over extremely long simulation times. We show that the spectral analysis provides us more quantitative information of this topological sharpening effect, especially as it differently impacts fluctuations of different wavelengths. We find substantial differences in the interfacial behavior of vertex and Voronoi models. In Voronoi models on long length scales there is an essentially constant degree of interfacial sharpening, independent of the imposed amount of microscopic surface energy or the temperture of our simulations. On short length scales, in constrast, the sharpening effect is extremely strong and strongly depends on the temperature and microscopic surface energy. We compare these predictions with a Brownian bridge model, suggesting that the interfacial effects we observe are due to a combination of a harmonic and a cusp-like potential with a varying population of four-fold vertices that changes the relative strength of these potentials. Unlike our results on the Voronoi model, we do not observe an interfacial sharpening effect in sufficiently equilibrated vertex models. Finally, we analyze existing experimental data \cite{guan2023interfacial}, which we find supports the role of a short length scale topological sharpening effect in real biological systems. 

\begin{figure}[t]
    \centering
    \includegraphics[width=0.95\columnwidth]{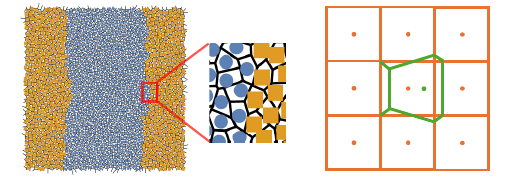}
    \includegraphics[width=0.99\columnwidth]{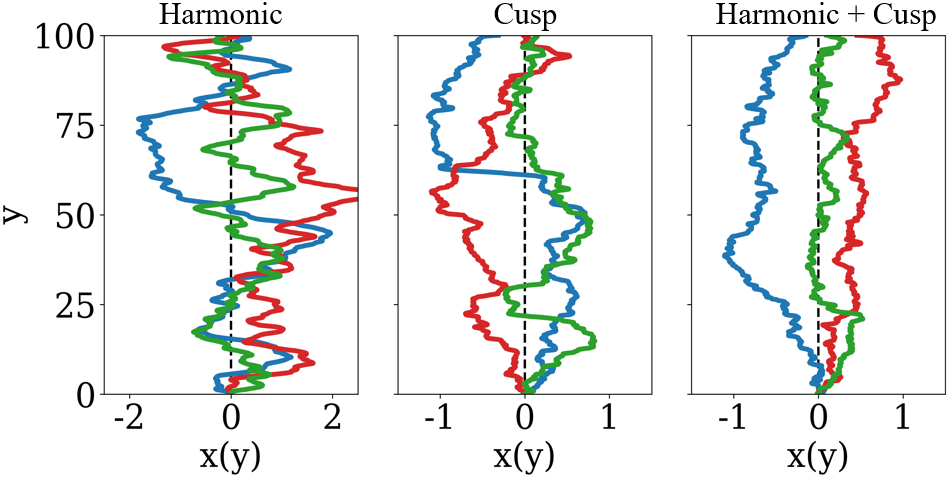}
    \caption{(Top left) Sample configuration of coexisting cell populations in the Voronoi model. (Top right) The breaking of four-fold vertices after a perturbation of the cell center illustrated in square lattices (orange: before perturbation; green: after perturbation). (Bottom) Sample trajectories for Brownian bridges in harmonic, cusp, and combined potentials. Brownian bridge trajectories were generated by Eq.~\ref{eq::defn_LandauRelaxationDynamics_numeric} using the free energy in Eq.~\ref{eq::HarmonicAndCusp_FreeEnergy} with $(\gamma,c)=(1.0,0.0)$, $(\gamma,c)=(0.0,1.0)$, and $(\gamma,c)=(1.0,1.0)$ in the left, center and right plots respectively, with a fixed value of $k_{B}T=0.5$.}
    \label{fig:methods}
\end{figure}

\section{Models and methods}
\subsection{Voronoi and vertex models}\label{sec:voroModel}
We simulate Voronoi and vertex models in a 2D square box with size $L$ and periodic boundary conditions. In the Voronoi model the cell shapes are determined by an instantaneous Voronoi tesselation of the current cell positions. In the vertex model the vertices themselves are the degrees of freedom, and cell shapes are traced out by connecting vertices around a cell. We initialize the system with strips of two different types of cells with width $L/2$, as shown in the inset of Fig.~\ref{fig:methods}. The forces on cells or vertices are calculated as the negative gradient of a cell-shape dependent energy \cite{sussman2017cellgpu, sussman2018soft}:
\begin{equation}
    \label{eq:agentBasedEnergy}
    E = \sum_{i=1}^N \left[ k_A (a_i -a_0)^2 +(p_i-p_0)^2 \right]+\sum_{\langle ij\rangle}\delta_{[i],[j]}\gamma_0 l_{ij}.
\end{equation}
In this energy, $N$ cells have  a quadratic energy penalty between instantaneous cell area $a_i$ and perimeter $p_i$ compared to the preffered area $a_0$ and preferred perimeter $p_0$. The second sum introduces an explicit interfacial tension over the edges $l_{ij}$ between neighboring cells with different types ($[i]$ means the type of cell $i$). Here and below we use $\gamma_0$ to denote the microscopic surface energy, and $\gamma$ to denote the effective value of the surface energy (e.g., as inferred from macroscopic observations of a simulation or experiment) We set the paramter $k_A$, which controls the relative area stiffness to perimeter stiffness, and the prefered area $a_0$ to unity. The preferred perimeter $p_0=4.0$ so that this confluent cellular systems are deep in the fluid phase \cite{bi2015density}. Unless otherwise specifed we perform our simulations with overdamped Brownian dynamics at temperature $T$: 
\begin{equation}
    \frac{d\vec{r}_i}{dt} = \vec{F}_i + 2\sqrt{k_BT/dt}\vec{\eta}, 
\end{equation}
where $\vec{\eta}$ is an uncorrelated gaussian noise of zero mean and unit variance. We use the cellGPU package (https://github.com/sussmanLab/cellGPU) for all simulations, and further details of the implementation of these equation of motion for Voronoi and vertex models are contained in Ref.~\cite{sussman2017cellgpu}.

\subsection{Interfacial width and interfacial spectrum}
The width of the interface between cell populations, $w$, is estimated by analyzing the density profile of one type of cell across the system. Given arrangements of cell populations as in  Fig.~\ref{fig:methods}, we average the density of a given cell type along the vertical direction and fit the resulting density profile, $\rho(x)$, to an error function~\cite{sides1999capillary}$\rho(x,w) = \frac{1}{2}\left(1+\text{erf}\left(\frac{x}{\sqrt{2}w}\right)\right)$. Here $x$ is the position relative to the mean location of the interface. The fluctuation spectrum of the interface $|h_k|$ can be obtained by applying a Discrete Fourier transform (DFT) to evenly spaced points on the interface $h(y)$, which are generated through shape-preserving piecewise cubic interpolation. The interpolation and DFT are implemented in MATLAB R2024a with the functions \textit{interp1} and \textit{fft}. The relation between the width and the spectrum is: 
\begin{equation} \label{equ:width}
    w^2=\sum_{k=-n}^n |h_k|^2 = |h_0|^2 + 2\sum_{k=1}^n |h_k|^2, 
\end{equation}
where $h_0 = \int h(x) dx$, as the average position of the interface, can be set to zero. We choose the number of points on the interface, $n$, to be equal to $L$ -- given that our unit of length is the square root of the average cell size, this is approximately the number of cells on the interface.

Based on the capillary wave theory (CWT) \cite{rowlinson2013molecular}, in the absence of external forces the spectrum of the thermal capillary waves is expected to be:
\begin{equation} \label{equ:CWTspectrum}
    \langle |h_k|^2 \rangle = \frac{k_BT}{\gamma L} \frac{1}{q^2}, 
\end{equation}
where $q=2\pi k/L$ and $\gamma$ is the effective surface tension. This can be derived with the energy equipartition theorem applied to the surface energy:
\begin{align}\label{equ:CWTenergy}
    E_s &= \gamma \int_0^L \left(\sqrt{1+h'(y)^2} -1\right) dy \nonumber \\
    &\approx \frac{\gamma}{2}\int_0^L h'(y)^2 dy  \\
    &= \gamma L \sum_{k=1}^n q^2 |h_k|^2, \nonumber
\end{align}
which gives $\gamma Lq^2|h_k|^2 = k_B T$ for each component of the spectrum. As a result, the average width can be obtained based on Eq.~\ref{equ:width} as:
\begin{equation}
    \langle w^2\rangle = \frac{k_BT L}{2\pi^2 \gamma}\sum_{k=1}^n \frac{1}{k^2} \approx \frac{k_B T L}{12\gamma}, 
    \label{equ:cwt_width}
\end{equation}
where the final approximate equality assumes $n\gg 1$. 

An important insight of earlier work\cite{sussman2018soft} is that in a Voronoi model the perturbation of a cell center near the interface may have a surface energy quite different from that of Eq.~\ref{equ:CWTenergy}. Near interfaces there is a tendancy for these models to spatially coordinate four-fold vertices along the interface (Fig.~\ref{fig:methods}), with cell displacements from these highly-registered configurations results in a discontinuous change in the set of interacting neighbors of the displaced cell. This suggests a surface energy that contains \emph{both} a standard harmonic contribution and a cusp-like term reflecting discontinuities in the topology of the cellular neighbor network:
\begin{equation}
    \epsilon_s = g_0 |h'(y)| + g_1 h'(y)^2.
\end{equation}
In the highly idealized geometry of square cells perfectly registered along the interface, the coefficients in front of the cusp and harmonic terms above were worked out as $g_0 \propto \gamma_0(\sqrt{2}-1)$ and $g_1 \propto \gamma_0(1-1/\sqrt{2})$. The precise prefactors are not relevant -- they will be sensitive to the density of four-fold vertices along the interface, and the geometry of cells near the interface -- but rather the  fact that $\gamma_0$ sets the scale of both the cusp-like and harmonic terms. 

The cusp-like potential gives a non-zero constant restoring forces for very small perturbations, stronger than the restoring forces proportional to the perturbations in the harmonic cases, which results in a sharpening effect\cite{sussman2018soft,lawson2024differences}. In order to understand the effect of the combined cusp-like and harmonic surface energy, we combine large-scale simulations along with a numerical analysis of a toy model of these interfaces, as decribed below. 

\subsection{Brownian Bridge Simulations\label{sec:BrownianBridgeSimulations}}
In our cell-based simulations (e.g. Fig.~\ref{fig:methods}) the position of the interface between cells of different type, $h(y)$, is constrained by the periodic boundary conditions of our simulation to have $h(0)=h(L)$, and the position of the interface in between the periodic boundaries reflects some sort of effective transverse confining potential for this interface. In this sense we can interpret each instance of $h(y)$ as the result of a ``Brownian bridge'' process \cite{ross1995stochastic}. Brownian bridges involve stochastic processes that are, e.g., constrained to both start and end at specific locations after a given amount of time. We thus interpret $y$ as the ``time'' over which a stochastic process evolves, simultaneously subject to a transverse restoring force of different functionals forms and the Brownian-bridge constraint that $h(0)=h(y)=0$. 

In the appendix we show analytically how this Brownian bridge framework reproduces the scaling of interfacial fluctuations predicted by the capillary wave theory described above. That is, we explicitly show that the width of interfaces generated by Brownian bridges subject to the surface energy of  Eq.~\ref{equ:CWTenergy} scales as $\langle\sigma^{2}\rangle\propto {k_{B}T}L/\gamma$. However, as discussed above, in the Voronoi model one expects that the interface lives not in a harmonic potential, but rather in a potential which combines harmonic \emph{and} a cusp-like terms which are each proportional to the microscopic surface energy $\gamma_0$. 

Including these cusp-like potentials makes analytic solutions to the Brownian bridge process more difficult, and so we turn to straightforward numerical implementations of this toy model of the cellular interface. We proceed as follows. First, we assume that local fluctuations of the interface away from $h^{\prime}=0$ obey Landau theory relaxation dynamics for a non-conserved order parameter field \cite{goldenfeld2018lectures}, we write
\begin{equation}
    \label{eq::defn_LandauRelaxationDynamics}
    \partial_{y}h^{\prime} = -\Gamma\frac{\delta{E_s}}{\delta{h}^{\prime}}+ \sigma\eta(y)\,,
\end{equation}
where $h^{\prime}=\partial{h}/\partial{y}$, $\Gamma$ is a constant of proportionality with units of inverse energy, and $\eta$ is a white noise process modulated by a strength $\sigma$. In order to impose the Brownian bridge constraint we modify the above equation to \cite{ross1995stochastic}
\begin{equation}
    \label{eq::defn_LandauRelaxationDynamics_numeric}
    \frac{dh^{\prime}(y)}{dy} = -\frac{-h(y)}{L-y} - \frac{\delta{E_s}[h^{\prime}(y)]}{\delta{h}^{\prime}} + \sigma\eta(y).
\end{equation}
The  $(-h/(L-y))$ term serves to enforce the Brownian bridge constraint $h(0)=h(L)$, and we work in units of $\Gamma=1$. In addition to the analytical derivation in the Appendix, we explicitly show in Fig.~\ref{fig:brownianBridge} that our numerical simulations of these equations using the CWT free energy in Eq.~\ref{equ:CWTenergy} reproduces all of the scalings of the CWT interfacial width result in Eq.~\ref{equ:cwt_width}. 

With this framework in place, we consider instead a cusp-like surface energy,
\begin{equation}
    \label{eq::Cusp_FreeEnergy}
    E_s[{h}^{\prime}(y)] = \int_{0}^{L}dy\,c\big|h^{\prime}\big|\, ,
\end{equation}
where $c$ is analogous to surface tension but has units of energy. As shown in Fig.~\ref{fig:brownianBridge}, we find that in this case that the average variance of a bridge scales as
\begin{equation}
    \label{eq::varianceCusp}
    \langle\sigma^{2}\rangle = \frac{k_{B}T}{\gamma_{\text{eff}}}L\,,
\end{equation}
where
\begin{equation}
    \label{eq:cuspSurfaceTension}
    \gamma_{\text{eff}} = \frac{c^{2}}{k_{B}T}
\end{equation}
is a temperature-dependent effective surface tension produced by the cusp potential.

\begin{figure}[t]
    \centering    \includegraphics[width=0.49\columnwidth]{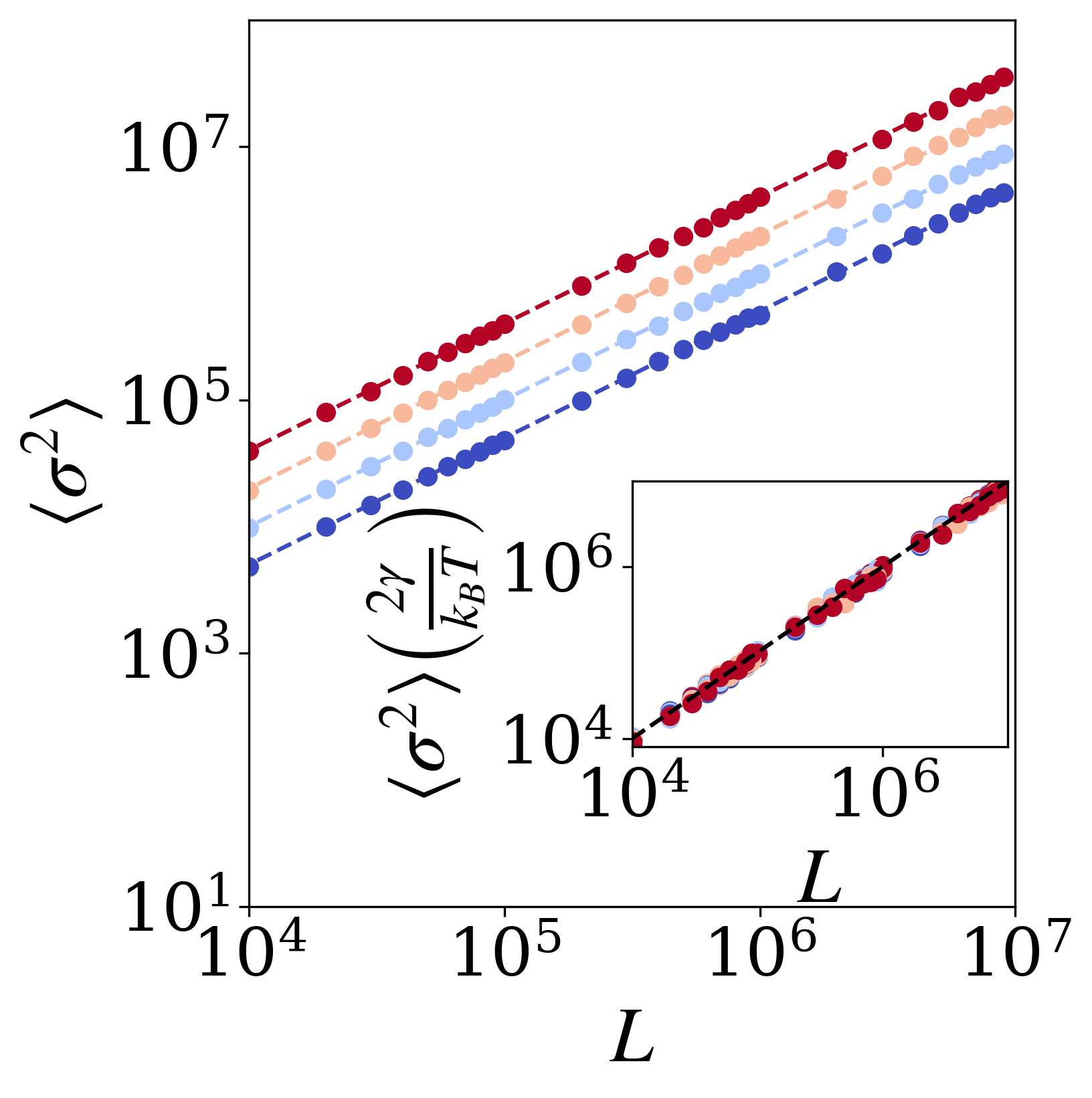}
    \includegraphics[width=0.49\columnwidth]{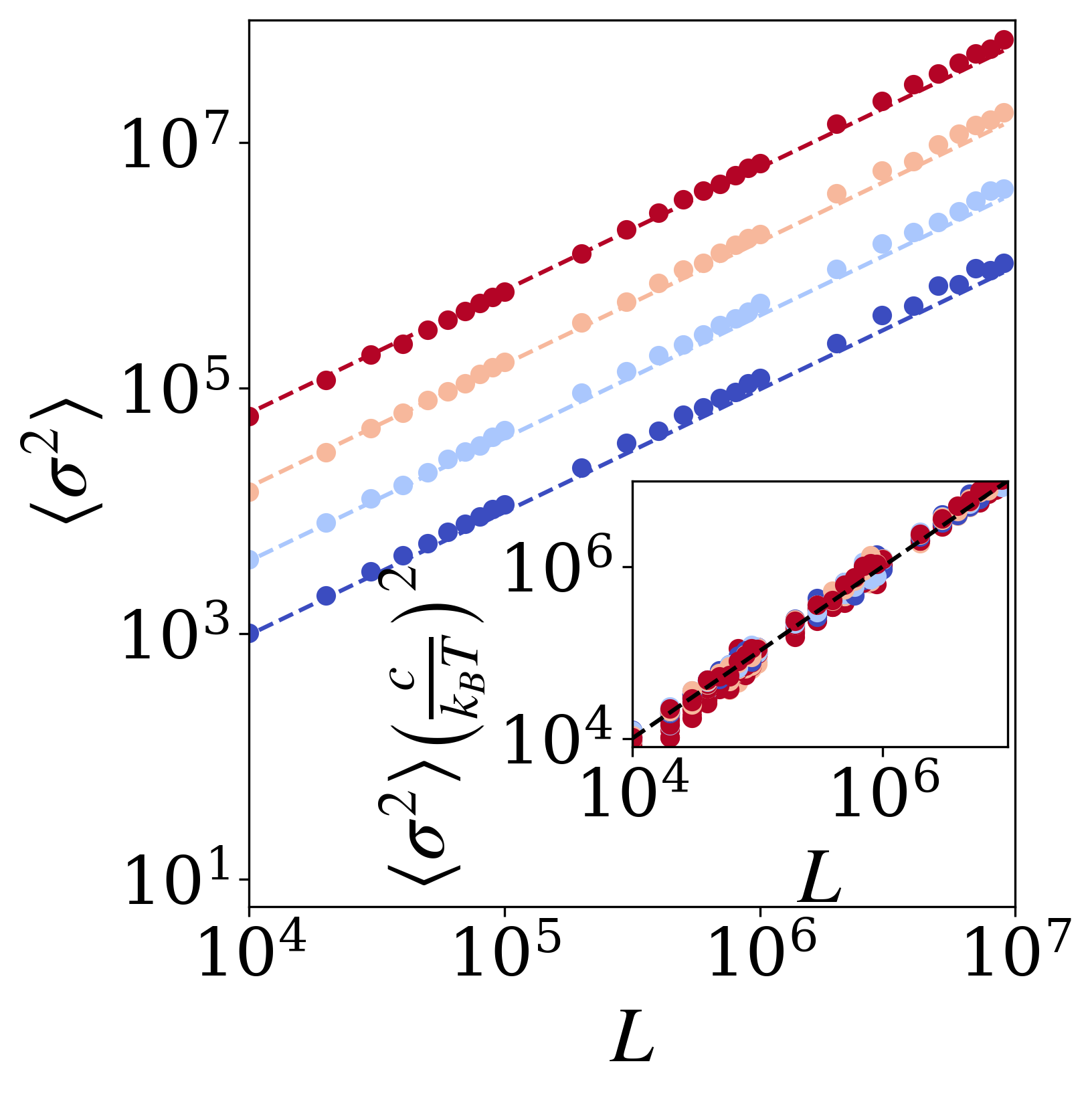}
    \caption{Brownian bridge variance scaling. The scaling of the average variance of interfaces $x(y)$ generated by numerical simulations of Eq.~\ref{eq::defn_LandauRelaxationDynamics_numeric} with a harmonic (left) and cusp (right) potential given by Eqs.~\ref{equ:CWTenergy} and \ref{eq::Cusp_FreeEnergy} respectively. Colors correspond to temperatures $k_{B}T\in\{0.005, 0.01, 0.02, 0.04\}$ (blue to red), and dashed lines in the main plot are fits to $\propto({k_BT}/\gamma)L$ and $\propto({k_{B}T}/c)^{2}L$. Insets in both figures demonstrate the expected collapse of the variance of these processes by the quantities indicated in the main text. Each contains data for $\gamma\in\{0.005, 0.01, 0.02, 0.04\}$ and $c\in\{0.004, 0.008, 0.016, 0.032\}$, and in each the dashed black line denoting the scaling $\propto{L}$.}
    \label{fig:brownianBridge}
\end{figure}

Finally, given the ``topological sharpening'' argument for the Voronoi model presented in Ref.~\cite{sussman2018soft} and outlined above, we extend this to the case of a surface energy with cusp-like and harmonic terms in the surface energy which are each proportional to the microscopic surface energy. As such, below when comparing with the results of the Voronoi model we present additional Brownian bridge simulations governed by an energy cost 
\begin{equation}
    \label{eq::HarmonicAndCusp_FreeEnergy}
    E_s[{h}^{\prime}(y)] = \int_{0}^{L}dy\, \left(c\big|h^{\prime}\big| + \frac{\gamma}{2}{h^{\prime}}^{2}\right),
\end{equation}
where both $c$ and $\gamma$ are set proportional to the microscopic surface energy $\gamma_0$. We find numerically that over a broad range of parameters the Brownian bridge ``interfaces'' generated by Eq.~\ref{eq::HarmonicAndCusp_FreeEnergy} have an effective surface tension which is approximately the sum of the harmonic and renormalized cusp surface tensions:
\begin{equation}
    \label{eq:harmonicAndCuspSurfaceTension}
    \gamma_{\text{eff}} \approx \frac{c^{2}}{k_{B}T} + 2\gamma.
\end{equation}

\section{Large-scale sharpening in Voronoi models}
\begin{figure}[t]
    \centering
    \includegraphics[width=0.9\columnwidth]{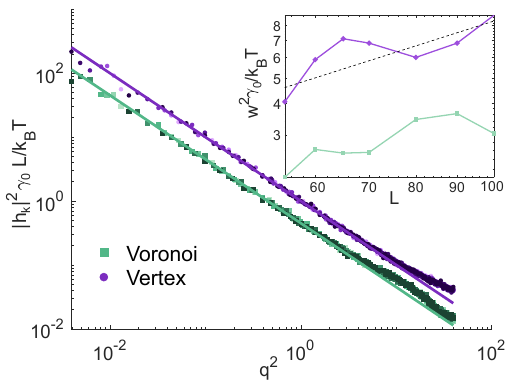}
    \caption{Large-scale interfacial sharpening. The interfacial spectrum is obtained for systems with sizes $L=55, 60, 65, 70, 80, 90, 100$ (colors from light to dark) and $\gamma_0=0.02$, each being average of 5 independent simulations. In each simulation, spectrum is calculated and averaged for data drawn at 50 evenly spaced time points after the system has reached equilibrium ($t>3\times 10^5 \tau$). After rescaling, the interfacial spectra for the vertex model and Voronoi models collapse onto separate curves. The solid lines are the fits of the low-wavevector data ($q^2 < 5$) to the expected form $ \frac{|h_k|^2\gamma_0 L}{k_B T} = \frac{A}{q^2}$ for $A\equiv \gamma_0/\gamma_{\text{eff}}$. $\gamma_0$ is the explicit surface tension in the models and $\gamma_{\text{eff}}$ is the effective surface tension. We find $A=0.998$ for vertex simulations and $A=0.450$ for Voronoi simulations. (inset) The rescaled interfacial width for systems of different sizes. The dashed line is the prediction of CWT.}
    \label{fig:large_scale}
\end{figure}

In order to compare the interfacial fluctuations in Voronoi and vertex models with CWT, we run simulations with system size $L$ ranging from $55$ to $100$ and explicit surface tension $\gamma_0=0.02$ for a long enough time to make sure the system has reached equilibrium. For each system, we measure the average width $w^2$, shown in the inset of Fig.~\ref{fig:large_scale}. We find that vertex simulations' results are in fact \emph{consistent} with CWT: at large length scales there is no obvious sharpening effect. In contrast, our Voronoi model simulations do show interfacial sharpening. Suprisingly, this sharpening effect ($w^2/w_0^2\approx 1/2$, with $w_0$ as the width predicted by CWT) is weak compared to the findings in previous literature \cite{sussman2018soft}. 

To begin resolving this finding, we calculate the full spectrum of interfacial fluctuations, as shown in Fig.~\ref{fig:large_scale}. We see that, except for the high-q regime (approximately $q^2>5$, which corresponds to sizes smaller than roughly three cells), the spectrum is consistent with the relation $|h_k|^2 \propto 1/q^2$. After rescaling, the spectra for systems of different sizes collapse onto one. By fitting the collapsed data with equation  $\frac{|h_k|^2\gamma_0 L}{k_B T} = \frac{A}{q^2}$, where $A\equiv \gamma_0/\gamma_{\text{eff}}$ is the sharpening factor, we get $A=0.998$ for vertex simulations and $A=0.450$ for Voronoi simulations. This is consistent with the conclusion we reached by directly measuring average width in the last paragraph, but fitting the spectrum allows a much more precise estimate of the sharpening compared to the much rougher estimate obtained from measuring interfacial widths. Given the much weaker effect we observe in the Voronoi model, it is perhaps understandable that the vertex model shows almost no effect at the relatively low surface tension used here, as the non-analytical behavior near four-fold vertices have been found not as strong as in Voronoi models \cite{lawson2024differences} (a subtletey compounded by the fact that most standard implementations of vertex models do not allow arbitrarily short edges between vertices; instead there is often a minimum edge length below which a T1 transition is triggered, which may in turn mask the effect of stable higher-order vertices \cite{yan2019multicellular}).

\begin{figure}[t]
    \centering
    \includegraphics[width=0.9\columnwidth]{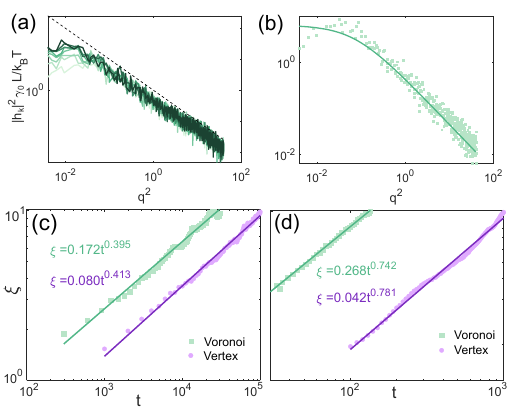}
    \caption{Dynamics of interfacial Spectrum (a) The interfacial spectra between $t=3000$ and $t=24000$ (from light to dark) are plotted Voronoi Brownian Dynamics simulations in systems with sizes $L=65, 70, 80, 90, 100$ and $\gamma=0.02$, each with 5 independent simulations. The dashed line is the theoretical prediction of classical CWT. (b) An example of fitting with equation \ref{equ:fitxi} for the interfacial spectrum at $t=2700$. (c-d) $\xi(t)$ for Voronoi and Vertex models with Brownian Dynamics (c) and Molecular Dynamics (d), fitted with equation (\ref{equ:fitxi}).}
    \label{fig:xi_time}
\end{figure}

Analyzing the full spectrum of interfacial fluctuations indeed helps address why we find much weaker sharpening effects in our Voronoi simulations. In Fig.~\ref{fig:xi_time}(a) we plot the spectrum of interfacial fluctuaions over time in our simulations. The magnitude of each fluctuation mode, $\langle|h_k|^2\rangle$, grows over time until reaching the equilibrium state, and naturally the small-$q$ modes take the longest time to equilibrate. Not only does this lead to spuriously low interfacial widths if the interfacial width is measured before the small-$q$ modes equilibrate, but the correlation time for the small-$q$ modes is also quite long. Thus, within a fixed-duration window of observation time, fewer time points can be used as independent data in the ensemble averaging. This is why in Fig.~\ref{fig:large_scale}, the data points for small $q$ modes are more noisy. In a linear regression as shown in Fig.~\ref{fig:large_scale}, the noisy data points in the low-q regime  do not influence the fitting result greatly. However, as the magnitude for the small $q$ modes dominate in the width, the interfacial width measurement are less precise, as revealed in the inset of Fig.~\ref{fig:large_scale}. We thus argue that spectral analyses are a preferred way to quantify the effective surface tension than directly measuring the interfacial width. We believe these issues affected previous measurements of interfacial sharpening in the Voronoi model \cite{sussman2018soft}, and also measurements of surface fluctuations in a variety of other soft-matter systems. It is worth checking whether the plateau in the spectrum within the low-q regime or sharpening effect for large-scale systems observed  in some previous studies reflects the long equilibration times hinted at above\cite{martinez2022morphological,del2019interface}.  

\begin{figure}[t]
    \centering
    \includegraphics[width=0.9\columnwidth]{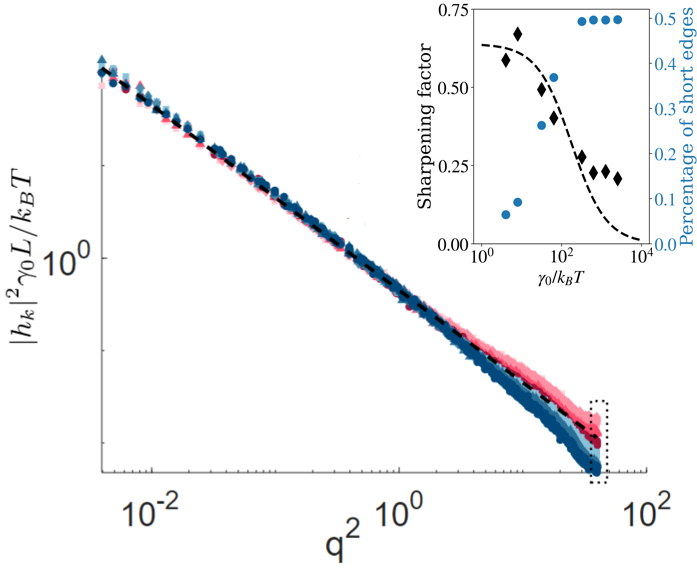}
    \caption{Interfacial Spectrum with different $\gamma_0$ and $k_B T$. All the red symbols are for $k_BT=0.005$ and the blue ones are for $k_BT=0.001$. For each temperature group, from light to dark, the $\gamma_0$ increases. Specifically, for $k_B T=0.005$, $\gamma_0=0.02,0.04,0.16,0.32$ respectively and for $k_B T=0.001$, $\gamma_0=0.32,0.64,1.28,2.56$ respectively. The black dashed line is for $\frac{\gamma_0 L}{k_BT}\langle|h_k|^2\rangle = A q^{-2}$ with $A=0.45$ obtained from the fitting in Fig.~\ref{fig:large_scale}. The original time step used in our simulation is $dt=0.01$, but for larger surface tension, to maintain numerical accuracy, we use smaller $dt$. Specifically, for $\gamma_0=0.64,1.28$, $dt=0.001$ is used and for $\gamma_0=2.56$, $dt=0.0001$ is used. (inset) The sharpening factor $A\equiv \gamma_0/\gamma_{\text{eff}}$ versus $\gamma_0/k_BT$ measured for high-q regime corresponding to about single-cell size (within the dotted box in the main figure) on the left y axis, and percentage of very short edges (smaller than 0.05) on the interface on the right y axis. Dashed black line denotes a fit to the the Brownian model prediction in Eq.~\ref{eq:brownianBridgeSharpeningFactorReduced} with $B_{1}=0.093$ and $B_{2}=1.565$.} 
    \label{fig:varyinggamma}
\end{figure}

We additionally point out that both our and many other measurements have been done for systems evolving according to Brownian dynamics. It is perhaps underappreciated that the approach to equilibrium under Brownian dynamics can be surprisingly slow, especially for the low-$q$ modes at the interface. To quantify this, we fit the spectra before equilibrium with the equation 
\begin{equation}
    \frac{\gamma L}{k_B T}\langle |h_k(t)|^2\rangle = \frac{A}{q^2+\xi(t)^{-2}}, 
    \label{equ:fitxi}
\end{equation}
where $\xi$ is defined as a characteristic length. When $\xi \gg q^{-1}$, Eq.~\ref{equ:fitxi} reduces to the equilibrium spectrum (Eq.~\ref{equ:CWTspectrum}). One example of the fitting is shown in Fig.~\ref{fig:xi_time}(b) and the fitted $\xi(t)$ for Voronoi and vertex simulations is shown in Fig.~\ref{fig:xi_time}(c). As a comparison, we implement the same process for simulations with Molecular Dynamics, which is suitable for regular liquids with momentum conservation, as shown in Fig.~\ref{fig:xi_time}(d). The $\xi(t)$ within a certain range, can be well fitted with power functions $\xi(t)=Bt^\alpha$, and the power for Brownian Dynamics is smaller (about a half) than that for Molecular Dynamics. For a system with size $L$ to fully equilibrate, it is required that $\xi \gg L/2\pi$ which means the characteristic time for equilibrium $\tau \propto L^{2.5}$ for Brownian Dynamics and $\tau \propto L^{1.25}$ for Molecular Dynamics. Because of this, although the simulation time in previous literature \cite{sussman2018soft} is long enough for the interface of a regular liquid system to reach equilibrium, the Brownian Dynamics simulations have not equilibrated so that the measured width (dominated by the low-q modes) is much smaller. This large-scale slow dynamics in systems with microscopic Brownian Dynamics, as also shown in a recent paper on coalescence \cite{yue2024coalescing}, again underscores the need for extra caution when studying the equilibrated state of highly dissipative biological systems. For example, in the advancing front of an epithelial monolayer, determining quasi-equilibrium is difficult without comparing the time scales of boundary equilibration and boundary movement. The time scale for boundary equilibration can be signifcantly increased by cell-substrate friction, particularly in large systems. Only when the front stops, such as due to contact inhibition when confronting another piece of epithelial monolayer, can measurements be confidently considered at equilibrium.  

The above results are for systems with relatively low surface tension. Then, we run more simulations by varying $\gamma_0$ and $k_BT$ to check how they influence the sharpening factor $A$, as shown in Fig.~\ref{fig:varyinggamma}. For Voronoi models, on large scales ($q>1$), after rescaling, all the spectra collapse onto one and the sharping factor $A$ is always around $0.5$. However, on small scales ($q<1$), the sharpening factor varies, which we will dig into in the next section.

\section{Small-scale sharpening in Voronoi models}
In contrast to the relatively weak but robust sharpening effects at the longest length scales, we continue to find very strong sharpening effects at smaller length scales. From Fig.~\ref{fig:varyinggamma}, in the regime $q>1$, corresponding to length scale of about a few cells, the magnitude of the modes decreases as we increase $\gamma_0/k_B T$ (and the effect grows strong at higher $q$). In the inset of Fig.~\ref{fig:varyinggamma} (left y axis), we show that the sharpening factor $A\equiv\gamma_0/\gamma_{\text{eff}}$ on the length scale of about one cell decreases from around $0.7$ to about $0.2$ when we increase $\gamma_0/k_BT$ from $4$ to $2560$. This change is consistent with the trend of increasing percentage of short edges on the interface, as shown on the right y axis in the inset of Fig.~\ref{fig:varyinggamma}, supporting our hypothesis that this small-scale sharpening is due to the four-fold vertices in the Voronoi model. To better understand how the cusp-like potential near the four-fold vertices influences the sharpening, we build a Brownian bridge model to compare with the Voronoi model simulations.

As discussed in Section \ref{sec:BrownianBridgeSimulations}, our Brownian bridge model predicts that an interface in the presence of a cusp potential should have a temperature-dependent \textit{effective surface tension} ($\gamma_{\text{eff}}$) given by Eq.~\ref{eq:harmonicAndCuspSurfaceTension}. The corresponding predicted sharpening factor is then given by
\begin{equation}
    \label{eq:brownianBridgeSharpeningFactor}
    \frac{\gamma_{0}}{\gamma_{\text{eff}}} = \frac{\gamma_{0}}{\frac{c^{2}}{{k}_{B}T}+2\gamma}\,.
\end{equation}
As noted above, both the cusp and harmonic terms have strength set by the microscopic surface energy. In the real cell systems we do not a priori know the relative density of four-fold vertices generating the cusps nor the precise geometry which sets the balance between the harmonic and cusp terms. Although they are in principle themselves dependent on the ratio $\gamma_0/(k_BT)$, for simplicity we simply adopt parameters $B_{1}$ and $B_{2}$ in the expression
\begin{equation}\label{eq:brownianBridgeSharpeningFactorReduced}
    \frac{\gamma_{0}}{\gamma_{\text{eff}}} = \frac{1}{B_{1}^{2}(\gamma_{0}/{k}_{B}T)+B_{2}}\,.
\end{equation}
as free but constant parameters quantifying these effects. 

We see that even with this simplifying assumption, the inset of Fig.~\ref{fig:varyinggamma} suggests that Eq.~\ref{eq:brownianBridgeSharpeningFactorReduced} provides a reasonable fit to the observed sharpening in the Voronoi model simulations over several orders of magnitude of $\gamma_{0}/{k}_{B}T$. In the limit $\gamma_{0}/{k}_{B}T\rightarrow\infty$ our Brownian bridge model predicts infinite sharpening as $\gamma_{\text{eff}}\rightarrow\infty$, whereas in Voronoi model simulations the sharpening appears to plateau at a non-zero value. The fact that the plateau in sharpening coincides with a plateau in the number of four-fold defects populating the interfaces suggests that the discrepancy with the Brownian bridge model is likely connected to non-negligible geometry-based variations of the parameters $B_{1}$ and $B_{2}$ in the $\gamma_{\text{eff}}\rightarrow\infty$ regime. Regardless, these results demonstrate that the small wavelength deviations from CWT scaling observed in Voronoi models can be understood as a consequence of the presence of four-fold vertices along the interface inducing a dependence of the effective surface tension on $\gamma_{0}/k_{B}T$.

\section{Discussion and outlook}

In this paper we investigated in detail the strength of interfacial fluctuations in models of cellular monolayers. Our focus has been on the quantification of how an unusual proposed ``topological sharpening'' effect at the microscopic scale renormalizes the effective surface tension at different length scales. Combining long-time simulations of Voronoi models with a spectrally resolved analysis of interfacial fluctuations, we quantitatively measured this sharpening effect on different lengths scales. We find that on large scales (more than $\sim$10 cells), this non-analytical local potential is renormalized to an equivalent harmonic potential with a modestly larger effective surface tension. In contrast, on small scales the interfacial fluctuations are strongly suppressed by the presence of  cusp-like contributions to the restoring  force at the interface. We verified that the broad features of this sharpening effect on small scales can be captured by a Brownian bridge model of the interface that combines cusp and harmonic contributions; the coefficients of this model would in general depend on both the fraction of four-fold vertices  at  the interface and the precise geometry cells adopt near it. 

The fact that there are short length scale deviations from the predictions of CWT is, of course, not in itself surprising: long lengthscale descriptions of physical systems always coarse grain over microscopic details, but at the scale of interacting particles these coarse grained descriptions discard relevant physical interactions. Importantly, we note that the \emph{trend} of this effect when studying interfacial fluctuations in particulate systems is typically in the opposite direction of what we find in these models of cellular interfaces: rather than a \emph{sharpening} effect, the finite size of the interacting particles lead to a roughening of the interface relative to CWT scaling \cite{willis2010thermal,shang2011fluctuating,del2019interface}. Similar roughening effects are also observed in vertex models (Fig.~\ref{fig:large_scale}) with small $\gamma_0$ when the topological effect of shape-based models is weak. 

\begin{figure}[t]
    \centering
    \includegraphics[width=0.9\columnwidth]{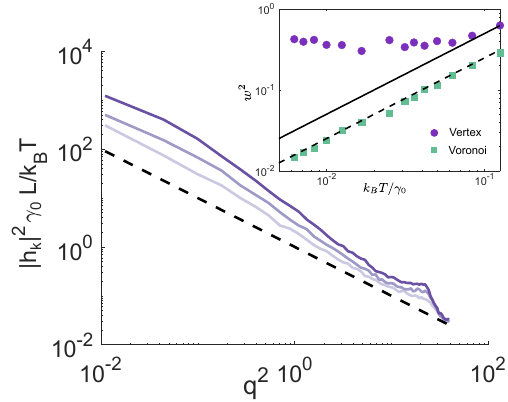}
    \caption{Interfacial width for Voronoi and vertex models with different $k_BT$ and $\gamma_0$. } 
    \label{fig:vertex}
\end{figure}

Interestingly, this behavior may be a point of difference between Voronoi and vertex models. The properties of both of these similar shape-based models are often quite similar, and although there are some subtle differences between them (for instance, it has been found that they have different athermal jamming transitions \cite{sussman2018no}), they are often regarded as largely interchangable when it comes to modeling. However, when we study their interfacial fluctuations we find qualitatively different behavior which is strongly sensitive to the imposed microscopic surface energy $\gamma_0$. For small values of $\gamma_0$, Fig.~\ref{fig:large_scale} shows that for sufficiently equilibrated vertex models clear sharpening effects do \emph{not} seem to be present. Furthermore, the large wavevector behavior shows a roughing effect that is similar to the typical behavior of particulate systems. This may be understandable, since in vertex models the perturbation of a single vertex does not  necessarily lead to topological changes around the whole cell of the sort shown for Voronoi tessellations in Fig.~\ref{fig:methods}. Previous studies have also indicated at most weak topological sharpening effects in vertex models by measuring the cusp-like restoring forces with small perturbations\cite{lawson2024differences}. 

The interfacial behavior of vertex models seems more complicated for large values of $\gamma_0$. The inset of Fig.~\ref{fig:vertex}) shows  a plateau in the interfacial width over a broad range of non-dimensionalized surface energies, where the magnitude of the plateau indicates an interface substantially rougher than that predicted by CWT. This is further reflected in the complicated changes in the spectrum seen in Fig.~\ref{fig:vertex}. We speculte that this roughening is, again, connected to the very different microscopic interfacial behaviors near four-fold vertices in Voronoi and vertex models observed in previous studies\cite{lawson2024differences}. The authors of that work found that the non-zero plateau of restoring forces due to the cusp-like potential have a very different dependence on explicit surface tension $\gamma_0$. A natural speculation is that this may also be connected to the introduction of short length scales that trigger topological transitions in common implementations of vertex models, but we have not found any direct evidence that varying this T1 length scale affects our findings. Further work may be needed to disentangle the subtle difference in interfacial behavior between Voronoi and vertex models.

Given the above discussion, an important question is whether topological sharpening effect can be seen in real biological systems. Previous studies have certainly observed the presence of four-fold vertices and of cell registration in epithelial tissues \cite{jacinto2000dynamic,lawson2024differences}, but this does not on its own imply that, e.g., four-fold vertices in cells play the same role that they do in highly simplified models of dense tissue. Measurements of interfacial width give an estimation of the \emph{effective} surface tension, but without knowning the microscopic details (such as precise measurements of cellular adhesion energies, cortical tensions, etc.), one cannot tell whether the effective surface tension does or does not match the microscopic one. We suggest that spectral analyses are a robust way to probe this question. 

\begin{figure}[t]
    \centering
    \includegraphics[width=0.9\columnwidth]{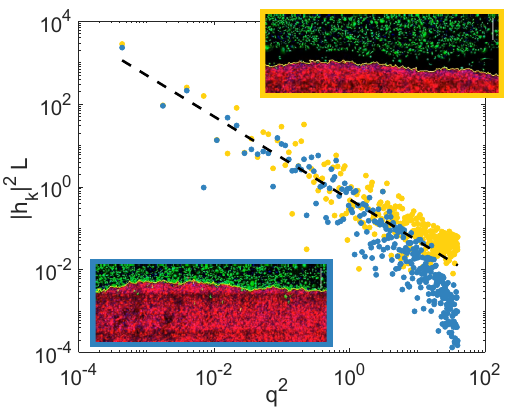}
    \caption{Interfacial spectrum calculated from experimental images (extracted from movie "Confrontation of MCF10A E-cad KD/MDA-MB-231 monolayers yields a straight interface" of Ref.~\cite{guan2023interfacial}). Yellow data points are estimated from a single snapshot at an early time (before the two types of epithelial cells contact). Blue data points are from a snapshot after a stable interface between the cell sheets has formed . The black dashed line is guide to the eye with the expected CWT scaling relation $|h_k|^2 L = 0.5/q^2$.} 
    \label{fig:exp}
\end{figure}

As an example, we analyzed video data of epithelial cell sheets presented as supplemental material in Ref. ~\cite{guan2023interfacial}, with estimates of two interfacial spectra shown in Fig.~\ref{fig:exp}. A relevant experiment in that paper involved the confrontation of two epithelial monolayers, composed of MCF10A-E-cad KD and MDA-MB-231 cells. At early times in the experiment -- before the two monolayers are in contact and, hence, before the possibility of forming any four-fold vertices -- we observe a spectrum of fluctuations consistent with CWT even at the highest wavevectors. In contrast, when the two cell sheets come into contact there is an obvious regime of high-$q$ sharpening, starting near $q\approx 1$. Fitting the low-$q$ regime of this data to the predictions of CWT, we find $k_BT/\gamma^y_{\text{eff}}\approx 0.538 (0.3277,0.7482)$ for the cells before the sheets contact and $k_BT/\gamma^b_{\text{eff}}\approx 0.3981 (0.2654, 0.5307)$ after the two cell sheets have formed a stable interface (where the numbers in parentheses represent a $95\%$ confidence bound). Thus, within the precision of our analysis of this existing experimental data, by looking at only the low-$q$ modes it is hard to determine whether there is an overall two-fold sharpening that our Voronoi model simulations would predict. On the other hand, the transition between the low-$q$ and high-$q$ regimes shows an effect quite like that in Fig.~\ref{fig:varyinggamma}, and supports the Voronoi model prediction of strong interfacial sharpening at small lengthscales. This reiterates the value of spectral analysis in the study of biological interfaces. 

We note that there are still some quantitative differences between our analysis of these experiments and our model simulations. Based on Fig.~\ref{fig:varyinggamma}, the high-$q$ regime sharpening becomes most  obvious when the effective surface tension as estimated from the low-$q$ regime is of order $\gamma_{\text{eff}}\approx 2\gamma_0 \approx 200 k_BT$. In Fig.~\ref{fig:exp}, the effective surface tension for blue points, which already show obvious high-q sharpening, is apparent even at our estimate of $\gamma_{\text{eff}}\approx 2.5k_BT$. As the obvious high-$q$ regime sharpening is enhanced when there are a large percentage of short edges (inset of Fig.~\ref{fig:varyinggamma}), the above inconsistency may suggests that in the experiments, the interfaces between two epithelial monolayers can have high-level of four-fold vertices, i.e. cell registration, even when the large-scale effective surface tension is not very strong. In the Voronoi model registration between cells across an interface is only enforced by the surface tension term, but of course in biological systems there may be other molecular mechanisms that support this.  For example, previous studies found that actin-based protrusions, like filopodia and lamellae, help with the correct matching of opposing cells along the fusion seam during dorsal closure \cite{jacinto2000dynamic,zhang2021mechanical}.

To further verify the above speculations, we propose the need for more experiments that combine large enough systems to provide a large range of the spectrum (as in Refs.~\cite{guan2023interfacial, heinrich2022self}), which also have clear images of cell membranes that can clearly show relative fractions of highly-coordinated vertices along the interface (as in Refs.~\cite{landsberg2009increased,iijima2020differential}). We believe that integrating modeling studies like those presented here with biological experiments involving genetic mutations of adhesion or actin cortex-related proteins is a promising route to enhancing our understanding of the specific roles these proteins play in the interfacial behaviors of epithelial tissues.

%\section*{Author Contributions}
%We strongly encourage authors to include author contributions and recommend using \href{https://casrai.org/credit/}{CRediT} for standardised contribution descriptions. Please refer to our general \href{https://www.rsc.org/journals-books-% databases/journal-authors-reviewers/author-responsibilities/}{author guidelines} for more information about authorship.

\section*{Conflicts of interest}
There are no conflicts to declare.

\section*{Acknowledgements}
We thank Mehran Kardar and Pierre Ronceray for motivating conversations. CRP was supported in part by Developmental Funds from the Winship Cancer Institute of Emory University.

\section{Appendix: Variance of Brownian bridge processes\label{sec:brownianBridgeAppendix}}
Here we derive the scaling relationship for the interfacial widths expected in our Brownian bridge framework. Brownian bridges $h(y)$ satisfy
\begin{equation}
    \label{eq::defn_BrownianBridge}
    {P}_{\text{bridge}}(h,y) = \frac{P\big(h=\delta\,,\,y=L\,|\,h_{0}=h\,,y_{0}=y\big)P\big(h,\,y\,|\,h_{0}=\delta\,,y_{0}=0\big)}{P\big(h=\delta\,,\,y=L\,|\,h_{0}=\delta\,,y_{0}=0\big)}
\end{equation}

As in the main text, we begin by assuming that local fluctuations of the interface away from $h^{\prime}=0$ obey Landau theory relaxation dynamics for a non-conserved order parameter field:
\begin{equation}
    \partial_{y}h^{\prime} = -\gamma\frac{\delta{E_s}}{\delta{h}^{\prime}}+ \sigma\eta(y)\, .
\end{equation}
As before, $\eta$ is a white noise process modulated by a strength $\sigma$ and $\gamma$ is a constant of proportionality with units of inverse energy (which we here take to be proportional to the harmonic surface energy term in Eq.~\ref{equ:CWTenergy}). In the smoothly varying (over-damped) limit, $|\partial_{y}h^{\prime}|\ll 1$, 
%Eqs.~\ref{eq::CWT_FreeEnergy}-\ref{eq::defn_LandauRelaxationDynamics}
Eqs.~\ref{equ:CWTenergy} and \ref{eq::defn_LandauRelaxationDynamics}
simplify to
\begin{equation}
    \label{eq::defn_RelaxationDynamics}
    \gamma\frac{dh}{dy} = \sigma\eta(y)\, ,
\end{equation}
where in thermal equilibrium
\begin{equation}
  \sigma \equiv \sqrt{2\gamma{k}_{B}T}.
\end{equation}

The free-particle Smoluchowski equation corresponding to Eq.~\ref{eq::defn_RelaxationDynamics} is given by
\begin{equation}
    \partial_{y}P(h,y\,|\,h_{0},y_{0}) = D\partial_{h}^{2}P(h,y\,|\,h_{0},y_{0})\,,
\end{equation}
where $D \equiv \frac{1}{2}\left(\sigma/(\gamma)\right)^{2}=(k_{B}T)/(\gamma)$, and admits the solution
\begin{equation}
    \label{eq::smoluchowskiSolution_freeParticle}
    P(h,y\,|\,h_{0},y_{0}) = \frac{e^{-\frac{(h-h_{0})^{2}}{4D(y-y_{0})}}}{\sqrt{4\pi{D}(y-y_{0})}}\,.
\end{equation}
Inserting Eq.~\ref{eq::smoluchowskiSolution_freeParticle} into Eq.~\ref{eq::defn_BrownianBridge}, then yields
\begin{equation}
    \label{eq::BrownianBridge_CWT}
    {P}_{\text{bridge}}^{\text{CWT}}(h,y) = \frac{1}{2}\sqrt{\frac{L}{\pi{D}{y}(L-y)}}e^{-\frac{Lh^{2}}{4Dy(L-y)}}\,.
\end{equation}
The variance of this distribution is computed straight-forwardly to be
\begin{equation}
    \label{eq::BrownianBridgeVariance_FreeParticle}
    \textrm{Var}(y) = 2D\frac{y(L-y)}{L}\,,
\end{equation}
which, averaged across the entire bridge, reduces to
\begin{equation}
    \label{eq:varianceHarmonic}
    \langle\textrm{Var}(y)\rangle = \frac{1}{3}\frac{k_{B}T}{\gamma}L\,.
\end{equation}
Although there is a difference in prefactor, we note that Eq.~\ref{eq:varianceHarmonic} yields precisely the temperature, length, and surface-tension scaling relation expected for a two-dimensional interface by standard CWT.

%%%END OF MAIN TEXT%%%


\balance


%%%REFERENCES%%%
\bibliography{SurfaceFluctuation} %You need to replace "rsc" on this line with the name of your .bib file

\providecommand*{\mcitethebibliography}{\thebibliography}
\csname @ifundefined\endcsname{endmcitethebibliography}
{\let\endmcitethebibliography\endthebibliography}{}
\begin{mcitethebibliography}{40}
\providecommand*{\natexlab}[1]{#1}
\providecommand*{\mciteSetBstSublistMode}[1]{}
\providecommand*{\mciteSetBstMaxWidthForm}[2]{}
\providecommand*{\mciteBstWouldAddEndPuncttrue}
  {\def\EndOfBibitem{\unskip.}}
\providecommand*{\mciteBstWouldAddEndPunctfalse}
  {\let\EndOfBibitem\relax}
\providecommand*{\mciteSetBstMidEndSepPunct}[3]{}
\providecommand*{\mciteSetBstSublistLabelBeginEnd}[3]{}
\providecommand*{\EndOfBibitem}{}
\mciteSetBstSublistMode{f}
\mciteSetBstMaxWidthForm{subitem}
{(\emph{\alph{mcitesubitemcount}})}
\mciteSetBstSublistLabelBeginEnd{\mcitemaxwidthsubitemform\space}
{\relax}{\relax}

\bibitem[Dahmann \emph{et~al.}(2011)Dahmann, Oates, and Brand]{dahmann2011boundary}
C.~Dahmann, A.~C. Oates and M.~Brand, \emph{Nature Reviews Genetics}, 2011, \textbf{12}, 43--55\relax
\mciteBstWouldAddEndPuncttrue
\mciteSetBstMidEndSepPunct{\mcitedefaultmidpunct}
{\mcitedefaultendpunct}{\mcitedefaultseppunct}\relax
\EndOfBibitem
\bibitem[Yang and Levine(2020)]{yang2020leader}
Y.~Yang and H.~Levine, \emph{Physical biology}, 2020, \textbf{17}, 046003\relax
\mciteBstWouldAddEndPuncttrue
\mciteSetBstMidEndSepPunct{\mcitedefaultmidpunct}
{\mcitedefaultendpunct}{\mcitedefaultseppunct}\relax
\EndOfBibitem
\bibitem[Pawlizak \emph{et~al.}(2015)Pawlizak, Fritsch, Grosser, Ahrens, Thalheim, Riedel, Kie{\ss}ling, Oswald, Zink, Manning,\emph{et~al.}]{pawlizak2015testing}
S.~Pawlizak, A.~W. Fritsch, S.~Grosser, D.~Ahrens, T.~Thalheim, S.~Riedel, T.~R. Kie{\ss}ling, L.~Oswald, M.~Zink, M.~L. Manning \emph{et~al.}, \emph{New Journal of Physics}, 2015, \textbf{17}, 083049\relax
\mciteBstWouldAddEndPuncttrue
\mciteSetBstMidEndSepPunct{\mcitedefaultmidpunct}
{\mcitedefaultendpunct}{\mcitedefaultseppunct}\relax
\EndOfBibitem
\bibitem[Yanagida \emph{et~al.}(2022)Yanagida, Corujo-Simon, Revell, Sahu, Stirparo, Aspalter, Winkel, Peters, De~Belly, Cassani,\emph{et~al.}]{yanagida2022cell}
A.~Yanagida, E.~Corujo-Simon, C.~K. Revell, P.~Sahu, G.~G. Stirparo, I.~M. Aspalter, A.~K. Winkel, R.~Peters, H.~De~Belly, D.~A. Cassani \emph{et~al.}, \emph{Cell}, 2022, \textbf{185}, 777--793\relax
\mciteBstWouldAddEndPuncttrue
\mciteSetBstMidEndSepPunct{\mcitedefaultmidpunct}
{\mcitedefaultendpunct}{\mcitedefaultseppunct}\relax
\EndOfBibitem
\bibitem[Steinberg(1962)]{steinberg1962mechanism}
M.~S. Steinberg, \emph{Proceedings of the National Academy of Sciences}, 1962, \textbf{48}, 1577--1582\relax
\mciteBstWouldAddEndPuncttrue
\mciteSetBstMidEndSepPunct{\mcitedefaultmidpunct}
{\mcitedefaultendpunct}{\mcitedefaultseppunct}\relax
\EndOfBibitem
\bibitem[Steinberg(2007)]{steinberg2007differential}
M.~S. Steinberg, \emph{Current opinion in genetics \& development}, 2007, \textbf{17}, 281--286\relax
\mciteBstWouldAddEndPuncttrue
\mciteSetBstMidEndSepPunct{\mcitedefaultmidpunct}
{\mcitedefaultendpunct}{\mcitedefaultseppunct}\relax
\EndOfBibitem
\bibitem[Amack and Manning(2012)]{amack2012knowing}
J.~D. Amack and M.~L. Manning, \emph{Science}, 2012, \textbf{338}, 212--215\relax
\mciteBstWouldAddEndPuncttrue
\mciteSetBstMidEndSepPunct{\mcitedefaultmidpunct}
{\mcitedefaultendpunct}{\mcitedefaultseppunct}\relax
\EndOfBibitem
\bibitem[Manning \emph{et~al.}(2010)Manning, Foty, Steinberg, and Schoetz]{manning2010coaction}
M.~L. Manning, R.~A. Foty, M.~S. Steinberg and E.-M. Schoetz, \emph{Proceedings of the National Academy of Sciences}, 2010, \textbf{107}, 12517--12522\relax
\mciteBstWouldAddEndPuncttrue
\mciteSetBstMidEndSepPunct{\mcitedefaultmidpunct}
{\mcitedefaultendpunct}{\mcitedefaultseppunct}\relax
\EndOfBibitem
\bibitem[Youssef \emph{et~al.}(2011)Youssef, Nurse, Freund, and Morgan]{youssef2011quantification}
J.~Youssef, A.~K. Nurse, L.~Freund and J.~R. Morgan, \emph{Proceedings of the National Academy of Sciences}, 2011, \textbf{108}, 6993--6998\relax
\mciteBstWouldAddEndPuncttrue
\mciteSetBstMidEndSepPunct{\mcitedefaultmidpunct}
{\mcitedefaultendpunct}{\mcitedefaultseppunct}\relax
\EndOfBibitem
\bibitem[Krieg \emph{et~al.}(2008)Krieg, Arboleda-Estudillo, Puech, K{\"a}fer, Graner, M{\"u}ller, and Heisenberg]{krieg2008tensile}
M.~Krieg, Y.~Arboleda-Estudillo, P.-H. Puech, J.~K{\"a}fer, F.~Graner, D.~M{\"u}ller and C.-P. Heisenberg, \emph{Nature cell biology}, 2008, \textbf{10}, 429--436\relax
\mciteBstWouldAddEndPuncttrue
\mciteSetBstMidEndSepPunct{\mcitedefaultmidpunct}
{\mcitedefaultendpunct}{\mcitedefaultseppunct}\relax
\EndOfBibitem
\bibitem[Brodland(2002)]{brodland2002differential}
G.~W. Brodland, \emph{J. Biomech. Eng.}, 2002, \textbf{124}, 188--197\relax
\mciteBstWouldAddEndPuncttrue
\mciteSetBstMidEndSepPunct{\mcitedefaultmidpunct}
{\mcitedefaultendpunct}{\mcitedefaultseppunct}\relax
\EndOfBibitem
\bibitem[Wiseman(1977)]{wiseman1977can}
L.~L. Wiseman, \emph{Developmental Biology}, 1977, \textbf{58}, 204--211\relax
\mciteBstWouldAddEndPuncttrue
\mciteSetBstMidEndSepPunct{\mcitedefaultmidpunct}
{\mcitedefaultendpunct}{\mcitedefaultseppunct}\relax
\EndOfBibitem
\bibitem[Ninomiya \emph{et~al.}(2012)Ninomiya, David, Damm, Fagotto, Niessen, and Winklbauer]{ninomiya2012cadherin}
H.~Ninomiya, R.~David, E.~W. Damm, F.~Fagotto, C.~M. Niessen and R.~Winklbauer, \emph{Journal of Cell Science}, 2012, \textbf{125}, 1877--1883\relax
\mciteBstWouldAddEndPuncttrue
\mciteSetBstMidEndSepPunct{\mcitedefaultmidpunct}
{\mcitedefaultendpunct}{\mcitedefaultseppunct}\relax
\EndOfBibitem
\bibitem[Monier \emph{et~al.}(2010)Monier, P{\'e}lissier-Monier, Brand, and Sanson]{monier2010actomyosin}
B.~Monier, A.~P{\'e}lissier-Monier, A.~H. Brand and B.~Sanson, \emph{Nature cell biology}, 2010, \textbf{12}, 60--65\relax
\mciteBstWouldAddEndPuncttrue
\mciteSetBstMidEndSepPunct{\mcitedefaultmidpunct}
{\mcitedefaultendpunct}{\mcitedefaultseppunct}\relax
\EndOfBibitem
\bibitem[Honda(1983)]{honda1983geometrical}
H.~Honda, \emph{International review of cytology}, 1983, \textbf{81}, 191--248\relax
\mciteBstWouldAddEndPuncttrue
\mciteSetBstMidEndSepPunct{\mcitedefaultmidpunct}
{\mcitedefaultendpunct}{\mcitedefaultseppunct}\relax
\EndOfBibitem
\bibitem[Bi \emph{et~al.}(2016)Bi, Yang, Marchetti, and Manning]{bi2016motility}
D.~Bi, X.~Yang, M.~C. Marchetti and M.~L. Manning, \emph{Physical Review X}, 2016, \textbf{6}, 021011\relax
\mciteBstWouldAddEndPuncttrue
\mciteSetBstMidEndSepPunct{\mcitedefaultmidpunct}
{\mcitedefaultendpunct}{\mcitedefaultseppunct}\relax
\EndOfBibitem
\bibitem[Alt \emph{et~al.}(2017)Alt, Ganguly, and Salbreux]{alt2017vertex}
S.~Alt, P.~Ganguly and G.~Salbreux, \emph{Philosophical Transactions of the Royal Society B: Biological Sciences}, 2017, \textbf{372}, 20150520\relax
\mciteBstWouldAddEndPuncttrue
\mciteSetBstMidEndSepPunct{\mcitedefaultmidpunct}
{\mcitedefaultendpunct}{\mcitedefaultseppunct}\relax
\EndOfBibitem
\bibitem[Sussman \emph{et~al.}(2018)Sussman, Schwarz, Marchetti, and Manning]{sussman2018soft}
D.~M. Sussman, J.~Schwarz, M.~C. Marchetti and M.~L. Manning, \emph{Physical review letters}, 2018, \textbf{120}, 058001\relax
\mciteBstWouldAddEndPuncttrue
\mciteSetBstMidEndSepPunct{\mcitedefaultmidpunct}
{\mcitedefaultendpunct}{\mcitedefaultseppunct}\relax
\EndOfBibitem
\bibitem[Sahu \emph{et~al.}(2021)Sahu, Schwarz, and Manning]{sahu2021geometric}
P.~Sahu, J.~Schwarz and M.~L. Manning, \emph{New Journal of Physics}, 2021, \textbf{23}, 093043\relax
\mciteBstWouldAddEndPuncttrue
\mciteSetBstMidEndSepPunct{\mcitedefaultmidpunct}
{\mcitedefaultendpunct}{\mcitedefaultseppunct}\relax
\EndOfBibitem
\bibitem[Lawson-Keister \emph{et~al.}(2024)Lawson-Keister, Zhang, Nazari, Fagotto, and Manning]{lawson2024differences}
E.~Lawson-Keister, T.~Zhang, F.~Nazari, F.~Fagotto and M.~L. Manning, \emph{PLOS Computational Biology}, 2024, \textbf{20}, e1011724\relax
\mciteBstWouldAddEndPuncttrue
\mciteSetBstMidEndSepPunct{\mcitedefaultmidpunct}
{\mcitedefaultendpunct}{\mcitedefaultseppunct}\relax
\EndOfBibitem
\bibitem[Kardar(2007)]{kardar2007statistical}
M.~Kardar, \emph{Statistical physics of fields}, Cambridge University Press, 2007\relax
\mciteBstWouldAddEndPuncttrue
\mciteSetBstMidEndSepPunct{\mcitedefaultmidpunct}
{\mcitedefaultendpunct}{\mcitedefaultseppunct}\relax
\EndOfBibitem
\bibitem[Guan \emph{et~al.}(2023)Guan, Lin, Chen, Lv, Li, and Feng]{guan2023interfacial}
L.-Y. Guan, S.-Z. Lin, P.-C. Chen, J.-Q. Lv, B.~Li and X.-Q. Feng, \emph{ACS nano}, 2023, \textbf{17}, 24668--24684\relax
\mciteBstWouldAddEndPuncttrue
\mciteSetBstMidEndSepPunct{\mcitedefaultmidpunct}
{\mcitedefaultendpunct}{\mcitedefaultseppunct}\relax
\EndOfBibitem
\bibitem[Sussman(2017)]{sussman2017cellgpu}
D.~M. Sussman, \emph{Computer Physics Communications}, 2017, \textbf{219}, 400--406\relax
\mciteBstWouldAddEndPuncttrue
\mciteSetBstMidEndSepPunct{\mcitedefaultmidpunct}
{\mcitedefaultendpunct}{\mcitedefaultseppunct}\relax
\EndOfBibitem
\bibitem[Bi \emph{et~al.}(2015)Bi, Lopez, Schwarz, and Manning]{bi2015density}
D.~Bi, J.~Lopez, J.~M. Schwarz and M.~L. Manning, \emph{Nature Physics}, 2015, \textbf{11}, 1074--1079\relax
\mciteBstWouldAddEndPuncttrue
\mciteSetBstMidEndSepPunct{\mcitedefaultmidpunct}
{\mcitedefaultendpunct}{\mcitedefaultseppunct}\relax
\EndOfBibitem
\bibitem[Sides \emph{et~al.}(1999)Sides, Grest, and Lacasse]{sides1999capillary}
S.~W. Sides, G.~S. Grest and M.-D. Lacasse, \emph{Physical Review E}, 1999, \textbf{60}, 6708\relax
\mciteBstWouldAddEndPuncttrue
\mciteSetBstMidEndSepPunct{\mcitedefaultmidpunct}
{\mcitedefaultendpunct}{\mcitedefaultseppunct}\relax
\EndOfBibitem
\bibitem[Rowlinson and Widom(2013)]{rowlinson2013molecular}
J.~S. Rowlinson and B.~Widom, \emph{Molecular theory of capillarity}, Courier Corporation, 2013\relax
\mciteBstWouldAddEndPuncttrue
\mciteSetBstMidEndSepPunct{\mcitedefaultmidpunct}
{\mcitedefaultendpunct}{\mcitedefaultseppunct}\relax
\EndOfBibitem
\bibitem[Ross(1995)]{ross1995stochastic}
S.~M. Ross, \emph{Stochastic processes}, John Wiley \& Sons, 1995\relax
\mciteBstWouldAddEndPuncttrue
\mciteSetBstMidEndSepPunct{\mcitedefaultmidpunct}
{\mcitedefaultendpunct}{\mcitedefaultseppunct}\relax
\EndOfBibitem
\bibitem[Goldenfeld(2018)]{goldenfeld2018lectures}
N.~Goldenfeld, \emph{Lectures on phase transitions and the renormalization group}, CRC Press, 2018\relax
\mciteBstWouldAddEndPuncttrue
\mciteSetBstMidEndSepPunct{\mcitedefaultmidpunct}
{\mcitedefaultendpunct}{\mcitedefaultseppunct}\relax
\EndOfBibitem
\bibitem[Yan and Bi(2019)]{yan2019multicellular}
L.~Yan and D.~Bi, \emph{Physical Review X}, 2019, \textbf{9}, 011029\relax
\mciteBstWouldAddEndPuncttrue
\mciteSetBstMidEndSepPunct{\mcitedefaultmidpunct}
{\mcitedefaultendpunct}{\mcitedefaultseppunct}\relax
\EndOfBibitem
\bibitem[Mart{\'\i}nez-Calvo \emph{et~al.}(2022)Mart{\'\i}nez-Calvo, Bhattacharjee, Bay, Luu, Hancock, Wingreen, and Datta]{martinez2022morphological}
A.~Mart{\'\i}nez-Calvo, T.~Bhattacharjee, R.~K. Bay, H.~N. Luu, A.~M. Hancock, N.~S. Wingreen and S.~S. Datta, \emph{Proceedings of the National Academy of Sciences}, 2022, \textbf{119}, e2208019119\relax
\mciteBstWouldAddEndPuncttrue
\mciteSetBstMidEndSepPunct{\mcitedefaultmidpunct}
{\mcitedefaultendpunct}{\mcitedefaultseppunct}\relax
\EndOfBibitem
\bibitem[Del~Junco and Vaikuntanathan(2019)]{del2019interface}
C.~Del~Junco and S.~Vaikuntanathan, \emph{The Journal of Chemical Physics}, 2019, \textbf{150}, 094708\relax
\mciteBstWouldAddEndPuncttrue
\mciteSetBstMidEndSepPunct{\mcitedefaultmidpunct}
{\mcitedefaultendpunct}{\mcitedefaultseppunct}\relax
\EndOfBibitem
\bibitem[Yue \emph{et~al.}(2024)Yue, Burton, and Sussman]{yue2024coalescing}
H.~Yue, J.~C. Burton and D.~M. Sussman, \emph{Physical Review Research}, 2024, \textbf{6}, 023115\relax
\mciteBstWouldAddEndPuncttrue
\mciteSetBstMidEndSepPunct{\mcitedefaultmidpunct}
{\mcitedefaultendpunct}{\mcitedefaultseppunct}\relax
\EndOfBibitem
\bibitem[Willis and Freund(2010)]{willis2010thermal}
A.~Willis and J.~Freund, \emph{Physics of fluids}, 2010, \textbf{22}, 022002\relax
\mciteBstWouldAddEndPuncttrue
\mciteSetBstMidEndSepPunct{\mcitedefaultmidpunct}
{\mcitedefaultendpunct}{\mcitedefaultseppunct}\relax
\EndOfBibitem
\bibitem[Shang \emph{et~al.}(2011)Shang, Voulgarakis, and Chu]{shang2011fluctuating}
B.~Z. Shang, N.~K. Voulgarakis and J.-W. Chu, \emph{The Journal of chemical physics}, 2011, \textbf{135}, 044111\relax
\mciteBstWouldAddEndPuncttrue
\mciteSetBstMidEndSepPunct{\mcitedefaultmidpunct}
{\mcitedefaultendpunct}{\mcitedefaultseppunct}\relax
\EndOfBibitem
\bibitem[Sussman and Merkel(2018)]{sussman2018no}
D.~M. Sussman and M.~Merkel, \emph{Soft matter}, 2018, \textbf{14}, 3397--3403\relax
\mciteBstWouldAddEndPuncttrue
\mciteSetBstMidEndSepPunct{\mcitedefaultmidpunct}
{\mcitedefaultendpunct}{\mcitedefaultseppunct}\relax
\EndOfBibitem
\bibitem[Jacinto \emph{et~al.}(2000)Jacinto, Wood, Balayo, Turmaine, Martinez-Arias, and Martin]{jacinto2000dynamic}
A.~Jacinto, W.~Wood, T.~Balayo, M.~Turmaine, A.~Martinez-Arias and P.~Martin, \emph{Current Biology}, 2000, \textbf{10}, 1420--1426\relax
\mciteBstWouldAddEndPuncttrue
\mciteSetBstMidEndSepPunct{\mcitedefaultmidpunct}
{\mcitedefaultendpunct}{\mcitedefaultseppunct}\relax
\EndOfBibitem
\bibitem[Zhang and Saunders(2021)]{zhang2021mechanical}
S.~Zhang and T.~Saunders, Seminars in Cell \& Developmental Biology, 2021, pp. 75--84\relax
\mciteBstWouldAddEndPuncttrue
\mciteSetBstMidEndSepPunct{\mcitedefaultmidpunct}
{\mcitedefaultendpunct}{\mcitedefaultseppunct}\relax
\EndOfBibitem
\bibitem[Heinrich \emph{et~al.}(2022)Heinrich, Alert, Wolf, Ko{\v{s}}mrlj, and Cohen]{heinrich2022self}
M.~A. Heinrich, R.~Alert, A.~E. Wolf, A.~Ko{\v{s}}mrlj and D.~J. Cohen, \emph{Nature communications}, 2022, \textbf{13}, 4026\relax
\mciteBstWouldAddEndPuncttrue
\mciteSetBstMidEndSepPunct{\mcitedefaultmidpunct}
{\mcitedefaultendpunct}{\mcitedefaultseppunct}\relax
\EndOfBibitem
\bibitem[Landsberg \emph{et~al.}(2009)Landsberg, Farhadifar, Ranft, Umetsu, Widmann, Bittig, Said, J{\"u}licher, and Dahmann]{landsberg2009increased}
K.~P. Landsberg, R.~Farhadifar, J.~Ranft, D.~Umetsu, T.~J. Widmann, T.~Bittig, A.~Said, F.~J{\"u}licher and C.~Dahmann, \emph{Current Biology}, 2009, \textbf{19}, 1950--1955\relax
\mciteBstWouldAddEndPuncttrue
\mciteSetBstMidEndSepPunct{\mcitedefaultmidpunct}
{\mcitedefaultendpunct}{\mcitedefaultseppunct}\relax
\EndOfBibitem
\bibitem[Iijima \emph{et~al.}(2020)Iijima, Sato, Kuranaga, and Umetsu]{iijima2020differential}
N.~Iijima, K.~Sato, E.~Kuranaga and D.~Umetsu, \emph{Nature Communications}, 2020, \textbf{11}, 6320\relax
\mciteBstWouldAddEndPuncttrue
\mciteSetBstMidEndSepPunct{\mcitedefaultmidpunct}
{\mcitedefaultendpunct}{\mcitedefaultseppunct}\relax
\EndOfBibitem
\end{mcitethebibliography}
\bibliographystyle{rsc} %the RSC's .bst file

\end{document}